%% file: main.tex
\documentclass[review]{elsarticle}

\usepackage{lineno,hyperref}
\modulolinenumbers[5]

\usepackage{enumerate}
\usepackage{paralist}
\usepackage{url}
\usepackage{xcolor}
\usepackage{natbib}
\usepackage{hyperref}
\usepackage{comment}
\usepackage{multirow}
\usepackage{subcaption}
\usepackage{caption}
\usepackage{amssymb}
\usepackage[inline, shortlabels]{enumitem}
\usepackage[russian,english]{babel}

\journal{Journal of Biomedical Informatics}

\begin{document}

\begin{frontmatter}

\title{Multimodal Model with Text and Drug Embeddings for Adverse Drug Reaction Classification}

\author[kfu,msu]{Andrey Sakhovskiy}
\ead{andrey.sakhovskiy@gmail.com}
\author[kfu,sber,hse]{Elena Tutubalina}
\ead{ElVTutubalina@kpfu.ru}

\address[kfu]{Kazan Federal University, 18 Kremlyovskaya street, Kazan, Russian Federation, 420008}
\address[msu]{Lomonosov Moscow State University, 1 Leninskie gory, Moscow, Russian Federation, 119991}
\address[sber]{Sber AI, 19 Vavilova St., Moscow, Russian Federation, 117997}
\address[hse]{National Research University Higher School of Economics, 11 Pokrovsky Bulvar, Moscow, Russian Federation, 109028}






\begin{abstract}
In this paper, we focus on the classification of tweets as sources of potential signals for adverse drug effects (ADEs) or drug reactions (ADRs). 
Following the intuition that text and drug structure representations are complementary, we introduce a multimodal model with two components. These components are state-of-the-art BERT-based models for language understanding and molecular property prediction. Experiments were carried out on multilingual benchmarks of the Social Media Mining for Health Research and Applications (\#SMM4H) initiative. 
Our models obtained state-of-the-art results of 0.61 F$_1$-measure and 0.57 F$_1$-measure on \#SMM4H 2021 Shared Tasks 1a and 2 in English and Russian, respectively. On the classification of French tweets from SMM4H 2020 Task 1, our approach pushes the state of the art by an absolute gain of 8\% F$_1$. Our experiments show that the molecular information obtained from neural networks is more beneficial for ADE classification than traditional molecular descriptors. 
The source code for our models is freely available at \url{https://github.com/Andoree/smm4h\_2021\_classification}.\\
\end{abstract}


\end{frontmatter}


\input{paper_main.tex}

\section*{Acknowledgements}
The work on novel multimodal models and manuscript has been supported by the Russian Science Foundation grant \# 18-11-00284. The work on baselines has been supported by a grant from the President of the Russian Federation for young scientists-candidates of science (MK-3193.2021.1.6).

\bibliography{ml}

\end{document}

%% file: paper_main.tex
\section{Introduction}
The popularity of social media as a source for health-related information has increased tremendously in the past decade.
One of the well-studied research areas is pharmacovigilance from social media data that focuses on discovering adverse drug effects (ADEs) from user-generated texts. ADEs\footnote{The terms adverse drug effects (ADEs) and adverse drug reactions (ADRs) are often used interchangeably.} are unwanted negative effects of a drug, in other words, harmful and undesired reactions due to its intake. 

In recent years, researchers have increasingly applied neural networks, including Bidirectional Encoder Representations from Transformers (BERT) \citep{devlin2018bert}, to ADE detection from texts \citep{klein2020overview,magge2021overview,miftahutdinov2020kfu,miftahutdinov2020biomedical}. This is directly related to the creation of annotated multilingual corpora and Social Media Mining for Health (\#SMM4H) shared tasks \citep{klein2020overview,magge2021overview}.
Given a tweet, participants of this shared task are required to detect whether the tweet contains a mention of ADE using natural language processing (NLP) techniques. However, these studies mostly share the same limitations: models consider textual information only without leveraging drug structure. In particular, \citep{wang2020islab,gusev2020bert} utilized transformer-based classifiers with ensemble modeling and undersampling ranking first on SMM4H 2020 task 1a and 1b, respectively. On the other hand, studies from cheminformatics and drug discovery areas have focused on the prediction of the side effects of a given drug \citep{dimitri2017drugclust,honda2019smiles,li2020trimnet,wang2019smiles}. These studies utilized supervised models trained and evaluated on a database of marketed drugs and ADRs the Side Effect Resource (SIDER) \citep{kuhn2016sider} from the MoleculeNet benchmark \citep{wu2018moleculenet}. Each chemical structure is encoded with molecular descriptors or neural representations that are usually fed to a multi-label classification model. In addition, a number of BERT architectures has been proposed for molecular property prediction such as SMILES-BERT \citep{wang2019smiles}, MolBERT \citep{fabian2020molecular}, ChemBERTa \citep{chithrananda2020chemberta} with fine-tuning on the SIDER dataset. 

Inspired by multimodal studies, we propose a novel method to utilize both textual and molecular information for ADE classification. We study the impact of using different molecular representation approaches, including traditional molecular descriptors calculated with Mordred \citep{moriwaki2018mordred} and BERT-based encoders ChemBERTa and MolBERT. We explore two strategies to fuse drug representations and tweet representations: straightforward concatenation of representations and use of a co-attention mechanism to integrate features of different modalities. Along with the textual information, the incorporation of molecular structure can aid in understanding the relationship between different pharmacological and chemical properties and the occurrence of ADEs. 

A preliminary version of this work has appeared in~\cite{sakhovskiy2021kfu}. Compared to the conference version, we have:
\begin{inparaenum}[(1)]
	\item significantly extended the experimental part of this work to assess the performance of the proposed multimodal model; in particular, we extended experiments to three datasets in English, French, and Russian. Our model achieved state-of-the-art results on SMM4H 2021 Tasks 1a \& 2 and SMM4H 2020 Task 1b; 
    \item extended the description of the experimental datasets for ADE classification;
	\item investigated model performance on different drug groups, adding new experimental results and conclusions;
	\item performed error analysis and discussed the limitations of our model.
\end{inparaenum}

In Section~\ref{sec:related} we present related work, Section~\ref{sec:data} describes the datasets, Section~\ref{sec:method} introduces our approach, Section~\ref{sec:results} describes and discusses experimental results, and Section~\ref{sec:conclusion} concludes the paper.


\section{Related Work}\label{sec:related}

\subsection{ADE classification} Although there is a wide range of supervised machine learning methods that have been applied to classify user-generated posts in English, a relatively small number of recent studies have been focused on texts in other languages, i.e., Russian and French.
In the SMM4H 2020 shared task~\cite{magge2021overview}, the precision of optimal ADE classification systems varies across different languages.
For English, precision previously has stayed in the range of 0.45-0.65 reaching a score of 0.64 with the winning system, while precision for Russian and French have stayed in the range of 0.34-0.54 and 0.15-0.33, respectively \citep{klein2020overview}. Most of these models leverage raw text to classify each text or a token in a text to ADE class. \cite{wang2020islab} pre-trained a large version of RoBERTa~\citep{liu2019roberta} on an unlabeled corpus of 6 million tweets in English where each tweet includes drug mention; they fine-tuned this model without any imbalance techniques on SMM4H data. \cite{gencoglu2020sentence} trained a fully-connected network on pre-computed embeddings obtained using a distilled version of multilingual sentence BERT to classify French tweets. \cite{gusev2020bert} used an ensemble of BERT-based models and logistic regression with undersampling to classify Russian tweets. \cite{miftahutdinov2020kfu} used an ensemble of ten EnRuDR-BERT models \citep{tutubalina2021russian} pretrained on 5M health-related user posts. Overall, the percentage of teams using Transformer~\cite{vaswani2017attention} architectures for ADE classification rose from 80\% in SMM4H 2020 to 100\% in SMM4H 2021 \citep{magge2021overview}.

Our model improves upon the state-of-the-art models in three very critical ways: (1) incorporation of drug representations, (2) extensive experiments on three languages, (iii) analysis of model performance for various Anatomical Therapeutic Chemical (ATC) main groups.

\subsection{Multimodal learning on biomedical tasks} A number of studies proposed multimodal deep learning models to combine information from multiple modalities. These models show promising results compared to uni-modal models on protein-protein and drug-drug interaction identification from scientific texts in English \citep{asada2018enhancing,saha2020amalgamation}, 
image captioning \citep{yu2019multimodal}, image ads understanding \citep{savchenko2020ad}. \cite{asada2018enhancing} utilize textual sentence representations and representation of the molecular structures of active substances for classification of drug-drug interactions (DDI). The authors proposed a network with two components. The first component of the architecture is a convolutional neural network, where the original text is converted into embedding representations using the word2vec model, which are then combined with the positional embeddings of the two drugs in the text. The second component is a graph convolutional network (GCN). Experiments show a 2.39\% increase in the F-measure on the DDI problem compared to methods that do not use graph representations of molecules as features. Similar to the DDI task, \cite{saha2020amalgamation} investigated multimodal networks on BioInfer and HRPD50 datasets for the protein-protein interaction (PPI) task. The authors used BioBERT pre-trained on large corpora of medical texts (PubMed, PMC) to obtain text representations. Spatial structure in PDB format or FASTA-sequence of each protein is available according to the corresponding identifiers (PDB ID, ensemble ID). To obtain features from FASTA sequences, the nucleotide sequences are first one-hot encoded and serve as the input for three convolutional layers. To obtain structural features, the coordinates of the atoms from the PDB file are converted into an adjacency matrix, and for each atom, its feature vector is calculated. The obtained information is used as the graph representation of the protein, which is fed to GCN. The authors used a transformer layer with attention combining three modalities; the output of this layer serves as an input to the softmax layer for the final classification. On both datasets, the authors achieve state-of-the-art results, in both cases improving the previous F-measure result by about 7\% in absolute terms.

\section{Data}\label{sec:data}
\input{data}

\section{Models}\label{sec:method}
In our work, we propose bi-modal ADE text classification models that combine textual and drug modalities.
More formally, consider we have a text $T$ that can be represented as a pair of textual modality $t$ and a drug modality that is a set of $k_T$ drug mentions:  $T = (t, D_T)$, where $D_T = (d_1, d_2,..., d_{k_T})$. To obtain two unimodal representations of T, we sample a random drug $d_i$ from $D_T$ and encode the $t$ and $d_{i}$ using two encoders: (i) a textual encoder $M_{text}$ and (ii) a drug encoder $M_{drug}$:
	$$ u^{T}_{text} = M_{text}(t)\,\,\,\,\,\, u^{T}_{drug} = M_{drug}(d_{i}) $$

	where $u^{T}_{text}\in\mathbb{R}^{d_t}$  is the embedding of textual modality $t$ and $u^{T}_{drug}\in\mathbb{R}^{d_d}$ is the embedding of drug $d_i$; $d_t$ and $d_d$ are the dimensionalities of the obtained uni-modal vector representations. The multimodal binary text classification task can be formulated as:
$$ f_{cl}(f_{mod}(u^{T}_{text}, u^{T}_{drug})) $$
where $f_{mod} : \mathbb{R}^{d_t + d_d}\rightarrow\mathbb{R}^{d_{bi}}$ is a modality combination function that provides a bi-modal representation of $T$, $f_{cl} : \mathbb{R}^{d_{bi}} \rightarrow \mathbb{R}$ is a fully connected classification network with sigmoid output activation. $d_{bi}$ is the dimensionality of the bi-modal vector representation.

We use the final embedding of the classification token ("[CLS]" token) as a textual representation of an input text. In the uni-modal approach, the textual representation $u^{T}_{text}$ is usually directly fed to a classifier, whereas for bi-modal models, we first combine the textual modality with the drug modality. We evaluate 2 ways to combine text and drug modalities: (i) concatenation of text and drug embedding; (ii) the scaled dot-product attention used in Transformer-based models~\cite{vaswani2017attention}. For the first way, we concatenate  the textual embedding and drug embedding:
$$f^{concat}_{mod}(u^{T}_{text}, u^{T}_{drug}) = [u^{T}_{text} \oplus u^{T}_{drug}]$$
Attention mechanism allows to learn a multimodal embedding of sample $T$ as a linear combination of two modalities $u^{T}_{text}$ and $u^{T}_{drug}$. For text classification, we use the sum of the multimodal embedding and the textual embedding:
$$f^{att}_{mod}(u^{T}_{text}, u^{T}_{drug}) = \alpha \cdot u^{T}_{text} + (1 - \alpha) \cdot u^{T}_{drug}$$
where $\alpha$ is the weight of textual modality obtained using attention mechanism.

Recently, a number of BERT-based models were proposed to learn flexible and high-quality molecular representations for drug discovery problems. These models are trained on string molecular representations in Simplified Molecular Input Line Entry System (SMILES)~\citep{weininger1988smiles} format to learn latent representations.

\subsection{Text encoders}
We used the following versions of BERT-based~\cite{devlin2018bert}:


\begin{itemize}
	\item For English data, we used the large version of RoBERTa~\citep{liu2019roberta}, pretrained on English texts of various domains. The total size of pretraining corpora is 160 GB. The model has 16 heads, 24 layers, 1024 embedding size, and a total of 355M parameters;
	\item For Russian data, we used EnRuDR-BERT\footnote{\url{https://huggingface.co/cimm-kzn/enrudr-bert}}, a bilingual model pretrained on the (i) RuDReC~\citep{tutubalina2021russian} corpus, a Russian corpus of drug reviews, and (ii) the English corpus, the same one used for pretraining of EnDR-BERT. The model has 12 heads, 12 layers, 768 embedding size, and a total of 110M parameters;
	\item For French data, we used CamemBERT$_{base}$~\citep{martin2020camembert},  a RoBERTa-based~\citep{liu2019roberta} model, pretrained on the French subcorpus of OSCAR corpus~\citep{suarez2019asynchronous}. The model has 12 heads, 12 layers, 768 embedding size, and a total of 110M parameters;
\end{itemize}

\subsection{Drug encoders}

To obtain drug information, we map drug names to their identifiers in DrugBank \citep{wishart2008drugbank} using the DrugBank terminology. The DrugBank database is a unique bioinformatics and cheminformatics resource that combines detailed drug (i.e., chemical, pharmacological, and pharmaceutical) data with comprehensive drug target information. We use Drugbank v. 5.1.8 that contains 14325 identifiers.

\subsubsection{ATC classification}

Nowadays, the Anatomical Therapeutic Chemical (ATC) classification system is the most widely recognized classification system for drugs. The World Health Organization (WHO) recommended the ATC classification system for the division of drugs into diverse groups according to the organ or system on which they act and/or their therapeutic and chemical characteristics. This system classifies drugs into 14 main groups\footnote{\url{http://www.whocc.no/atc/structure_and_principles/}}: (A) alimentary tract and metabolism; (B) blood and blood-forming organs; (C) cardiovascular system; (D) dermatologicals; (G) genitourinary system and sex hormones; (H) systemic hormonal preparations, excluding sex hormones and insulins; (J) anti-infectives for systemic use; (L) antineoplastic and immunomodulating agents; (M) musculoskeletal system; (N) nervous system; (P) antiparasitic products, insecticides, and repellents; (R) respiratory system; (S) sensory organs; (V) various. \cite{chen2012predicting} used chemical-chemical interactions and similarities to predict ATC codes of drugs since compounds with similar physicochemical properties are often involved in similar biological activities. Following this assumption, we encode a drug as a sparse 14-dimensional vector with zero values for ATC codes to which a drug does not belong. We adopt information about ATC main groups from Drugbank for each drug.

\subsubsection{Molecular descriptors}
Similar molecular structures have similar molecular properties. A molecular descriptor is defined as the ``final result of a logical and mathematical procedure, which transforms chemical information encoded within a symbolic representation of a molecule into a useful number or the result of some standardized experiment" \citep{consonni2010molecular}. We use molecular-descriptor calculation library Mordred \citep{moriwaki2018mordred} to calculate 2 thousand descriptors for each drug from Drugbank.

\subsubsection{MolBERT}
MolBERT uses two domain-specific training objectives to improve on BERT: SMILES equivalence and calculated molecular descriptor prediction \cite{fabian2020molecular}. They also used traditional masked language modeling. The SMILES equivalence task is to predict whether the two SMILES represent the same molecule based on the cross-entropy loss. The descriptor prediction task is to predict the set of descriptors for each molecule based on the mean squared error over all predicted values and scores from RDKit. The final loss is given by the arithmetic mean of all task losses. The authors used 1.5M compounds for pre-training. We use the official checkpoint with 12 attention heads and 12 layers trained for 100 epochs and available at \urlstyle{tt}\url{https://github.com/BenevolentAI/MolBERT}.

\subsubsection{ChemBERTa}
 ChemBERTa follows the RoBERTa architecture with 12 attention heads and 6 layers \cite{chithrananda2020chemberta}. The authors adopted a  pretraining procedure from RoBERTa, which masks 15\% of the tokens in each input string. They used a dataset of 77M unique SMILES for pre-training. The official checkpoint are available at \urlstyle{tt}\url{https://huggingface.co/seyonec/ChemBERTa_zinc250k_v2_40k}.

\subsection{Classification network}

For classification, we utilize a fully-connected classification network. The classifier includes one hidden layer, GeLU~\cite{hendrycks2016gaussian} activations and 0.3\% dropout. We use sigmoid as an output layer activation and binary cross-entropy as the loss function.  We freeze the first 5 Transformer encoder layers and the embedding layer of the encoder that shows higher results on the dev set.

\section{Experiments}\label{sec:results}

\subsection{Experimental setup}

We compared the performance of our proposed bi-modal models with three uni-modal (text-only) baselines:
\begin{itemize}
	\item Support Vector Machines (SVM) \citep{boser1992training} classifier over TF-IDF vectors. Before TF-DF vectorization, we tokenized and lemmatized each text using Spacy\footnote{\url{https://spacy.io}} and filtered out stop words using language-specific lists of stopwords obtained from NLTK \citep{bird2009natural}. We used TF-IDF and SVM implementations from Scikit-learn \citep{scikit-learn}.
	\item Convolutional Neural Network (CNN) \citep{lecun1998gradient, kim2014convolutional} over pretrained Word2vec \citep{word2vec2013efficient} or Fasttext \citep{bojanowski2017enriching} word embeddings. We tokenized each text using Spacy and filtered out stop words using stopwords lists obtained from NLTK. Domain-specific biomedical English Word2Vec\footnote{\url{https://bio.nlplab.org}} and Russian FastText\footnote{\url{https://github.com/cimm-kzn/RuDReC}} models and a general-domain French FastText\footnote{\url{https://fasttext.cc/docs/en/crawl-vectors.html}} model were used to obtain word embeddings for the CNN baseline. Each CNN model was composed of three parallel CNN layers followed by Rectified Linear Unit (ReLU) activation and one-dimensional max-pooling. A final text representation was obtained as a concatenation of these three resulting representations fed to a fully-connected layer. Each of the three CNN layer had 128 filters with window sizes of 3, 4, and 5. We trained each CNN model for 15 epochs with a learning rate of $1 \cdot 10^{-4}$ and made a prediction using model parameters from the best epoch in terms of validation F1-score.
	\item BERT-based unimodal classifier.
\end{itemize}


For Russian and French datasets, we trained each BERT-based model for 10 epochs with a batch size of 64 and a learning rate of $3 \cdot 10^{-5}$ using Adam optimizer~\cite{DBLP:journals/corr/KingmaB14}. For test set predictions, we used model weights from the best epoch in terms of dev set F1-score. For English data, we trained each RoBERTa-based model for 5 epochs with the learning rate of $1 \cdot 10^{-5}$ and used the model weights from the last training epoch to make a text set prediction. Since English and French training sets have a strong class imbalance, we applied a positive class over-sampling to English and French corpora.




\input{tables}

CNN baselines, unimodal and bi-modal BERT-based models were implemented using the PyTorch library \citep{NEURIPS2019_9015}. Pretrained BERT-based models were obtained using the Transformers library \citep{wolf-etal-2020-transformers}. For each dataset and training setup, we trained 10 BERT-based models with different initializations of classifier weights and calculated the mean quality and standard deviation of F\textsubscript{1}-scores (F\textsubscript{1}), precision (P), and recall (R) for the \textit{ADE} class.

\subsection{ADE text classification}

Table~\ref{tab:smm4h_20_fr_results} presents the results on the French set for SMM4H 2020 Task 1b. Based on the results, we can draw the following conclusions. First, the highest F\textsubscript{1}-scores were achieved by bi-modal models with concatenated modalities. In particular, the bi-modal model with MolBERT drug embedding outperformed the best official results of the SMM4H 2020 competition and text-only model by 8.5\% in terms of F\textsubscript{1}-score, achieving a new state-of-the-art. Second, the performance of models with cross-attention is significantly lower than the ones with modalities concatenation. However, cross-attention models with chemBERTa and MolBERT embeddings slightly outperformed the text-only model by 1.8\% and 0.9\%. Finally, uni-modal and bi-modal BERT-based models significantly outperform unimodal SVM and CNN baselines, thus proving the superiority of complex neural networks over simple neural models and traditional machine learning algorithms in the task of ADE text classification. 

Table~\ref{tab:smm4h_21_ru_results} shows the performance of uni-modal and bi-modal models on the Russian tweet corpus of SMM4H 2021. According to the results, we can make the following observations. First, the two best-performing models in terms of F\textsubscript{1}-score are bi-modal models that utilize MolBERT drug embeddings. Concatenation-based model slightly outperforms attention-based model (0.3\%) and unimodal BERT (0.8\%). Second, bi-modal models with concatenated modalities outperform the ones with cross-attention. Third, our final SMM4H 2021 submission combined the results of ten multimodal models with the same settings using a simple voting scheme with the intent of increasing the robustness of the final system. As shown in Table \ref{tab:smm4h_21_ru_results_magge}, this model achieved state-of-the-art results showing 4\% improvement in terms of F$_{1}$-measure over a single model.

Table~\ref{tab:smm4h_21_en_results} shows the performance of uni-modal and bi-modal models on English tweets.
Since there are strict limits with 3 submissions per day for the official test set on the SMM4H 2021 Codalab page\footnote{\url{https://competitions.codalab.org/competitions/28766}}, we use 20\% of the original train set as a test set for the evaluation of our models. Several observations can be made based on Table~\ref{tab:smm4h_21_en_results}.
First, for the dev set, the majority of bi-modal models do not show a significant performance difference compared to the uni-modal classifier, but the highest F\textsubscript{1}-score is achieved by the descriptors-based model with cross-attention that outperforms the text-only model by 0.6\% in terms of F\textsubscript{1}.
Second, the best F\textsubscript{1} on the our test set is significantly higher than on the official test set (0.799 vs 0.61).
Third, as shown in Table \ref{tab:smm4h_21_ru_results_magge}, our official SMM4H 2021 submission combined the results of ten models with the same settings using a simple voting scheme achieved the results on par with the best-performing team \citep{ramesh2021bert} that utilized the RoBERTa model with undersampling and oversampling. 

\subsubsection{Performance on ATC-stratified test set} 

In this subsection, we investigate if bi-modal models perform better in a case when train and test sets are similar in terms of covered ATC classes. For this purpose, we compared the proposed bi-modal models and unimodal BERT on  ATC-stratified SMM4H 2021 English data split, i.e., on a dataset that has the same distribution of ATC groups among train and test sets. Table~\ref{tab:smm4h_21_stratified_en_results} presents the results. The following observations can be made. Bi-modal models that are based on ATC groups, ChemBERTa embeddings, and chemical descriptors consistently perform no worse than the unimodal BERT in terms of F1-score and have lower performance variance. The highest quality is observed for concatenation-based ChemBERTa and descriptors models that outperform unimodal BERT by 1.1\% and 1.0\% F1, respectively. But the difference between bi-modal and unimodal models is statistically significant ($\rho \leq 0.05$) for the descriptors-based model with modalities concatenation only. Based on the results, we cannot conclude there exists a dependency between the similarity of train and test sets' drug groups distribution and the quality of a bi-modal model.

\subsubsection{Performance on different drug groups} 
\input{tables2}

In this subsection, we explore how the incorporation of additional modality affects the classification performance on tweets that mention drugs of different therapeutic groups. For the Russian and English corpus of SMM4H 2021, we took top-performing bi-modal models in terms of F\textsubscript{1}-score and compared their performance with uni-modal models on test set subsets of 5 most frequent ATC groups. 

Table~\ref{tab:quality_by_group} presents the results for the English and Russian corpora of SMM4H 2021. Based on the results, we can draw the following conclusions. First, bi-modal models show an increase of more than 1\% F\textsubscript{1} compared to uni-modal models for the most frequent ATC group ``Nervous system`` on both datasets. Second, for each dataset and therapeutic group, bi-modal models obtain higher precision. The precision growth varies from 0.7\% (``Dermatologicals``) to 3.8\% (``Sensory organs``) for the Russian corpus and from 1.3\% (``Dermatologicals``) to 5.1\% (``Nervous system``) for English corpus. Third, for all top-5 therapeutic groups of Russian corpus, bi-modal models achieved higher F\textsubscript{1}. The highest F\textsubscript{1} increase is achieved for group ``Sensory organs`` (+3.3\%), and the lowest increase of 0.1\% is insignificant and is achieved for group ``Anti-infectives for systemic use``. For English data, the use of bi-modal results in F\textsubscript{1} decrease for groups ``Alimentary   tract and metabolism`` (-3.0\%) and ``Sensory organs`` (-3.7\%). Finally, for a fixed therapeutic group, the sign of performance difference between bi-modal and uni-modal models varies for different languages. In particular, for group ``Sensory organs`` the bi-modal model shows a 3.3\% growth and 3.7\% decrease of F\textsubscript{1} for Russian and English corpus, respectively. Hence, we can conclude that the performance of the bi-modal classification approach is not dependent on only ATC groups' distribution of input data.


\subsection{Error analysis}

\begin{otherlanguage}{russian} 

\begin{table*}[!t]
\centering
\caption{SMM4H 2021 data samples with predictions by our model. The "Error type"\ column presents prediction error type according to the human annotation: "False positive"\ stands for the cases when the model incorrectly labels a text sample as ADE, and "False negative"\ stands for the cases when the model incorrectly labels a text sample as non-ADE. \label{tab:error_analysis}} 
\begin{tabular}{|p{1.75cm}|p{7.5cm}|p{1.25cm}|}
\hline
\textbf{Error reason}& \textbf{Text sample} & \textbf{Error type} \\ \hline
Positive effect & however the prozac is making her more talkative than usual (if also hilarious and lucid) so i don't think she's fully listening to my ideas & False Positive \\ \hline
Positive effect & Nicotine always makes me get out of a bad mood & False Positive \\ \hline
Positive effect & Hyper af! I feel like a bird on vyvanse & False Positive \\ \hline
Positive effect & не знаю от чего меня так вштырило от ношпы или парацетамола но я не могу перестать танцевать и мне в общем то хорошо (I don't know if Drotaverine or Paracetamol got me this feeling, but I feel pretty good and can't stop dancing)  & False Positive \\ \hline
Annotation error & Yay! Cymbalta is making me numb again. & False Positive \\ \hline
Annotation error & alright time to find out if a ton of vyvanse and not sleeping for 2 days can get a trimester worth of homework done for 2 classes & False Negative  \\ \hline
Ambiguous sample & half a trazodone had me sleeping heavy this morning.... & False positive \\ \hline
Ambiguous sample & Последствия наложения парацетамола на моё организм: проснулся с температурой 35. (The consequences of applying paracetamol on my body: I woke up with a temperature of 35.0) & False positive \\ \hline
\end{tabular}
\end{table*}

\end{otherlanguage} 

Table~\ref{tab:error_analysis} shows error analysis for sample texts predicted by our model. First, the samples that mention an unexpected positive drug effect are often wrongly labeled as ADE samples. It seems like the modern BERT-based models fail to distinguish the beneficial drug effects from the unpleasant ones. Second, the SMM4H corpora are noisy and contain annotation errors. Finally, there are ambiguous samples that mention a drug effect but do not specify if it's a desired effect or an adverse one. The existing text-only classification models are bound to fail on such samples as the textual modality does not provide enough information for proper classification. Thus, the models have to utilize external knowledge that can be presented in drug metadata explicitly or implicitly encoded in a drug representation.

\section{Discussion}\label{ssec:discuss}

In our work, we propose a multimodal model based on text and drug representations. Our experiments on three datasets demonstrate that the proposed model outperforms models based on text representations only. We have shown how our model improves upon the current state of the art and analyzed the influence of training set size on the results.

First, we investigated monolingual models only. These models still require manually annotated training data, which is not available for many languages. We can investigate how to transfer knowledge from one language to another. We believe that drug representations can help to gain improvement in this setup, since multilingual representations are not perfectly aligned, especially, for domain-specific topics such as biomedicine. In contrast, drug representations will be the same and can serve as implicit knowledge anchors about differences in drug groups. 

Second, we can encode additional knowledge into drug representations by training multi-label classifiers for the prediction of the side effects of a given drug on the SIDER dataset. Another direction is multimodal multitask learning on textual datasets and the SIDER dataset that will require jointly modeling the drug and the text modalities. We expect that further improvements may be possible with more research investigating other networks and developing a better joint loss function for two tasks.

Third, zero-shot cross-drug class evaluation is still an open research direction. A recurring limitation, which arises with supervised models, is bias in training data; annotated corpora include texts about a small number of drugs (at best about dozens) while there are over 20,000 prescription drug products approved for marketing\footnote{https://www.fda.gov/about-fda/fda-basics/fact-sheet-fda-glance}. The model performance is evaluated under the implicit hypothesis that the training data (source) and the test data (target) come from the same underlying distribution (i.e., both sets include drugs from the same ATC main group). It would be interesting to rebuild train/dev/test sets to model cross-drug class evaluation.

\section{Conclusion}\label{sec:conclusion}
In this paper, we studied the task of discovering adverse drug effects (ADEs) presence in user-generated tweets about drugs. In this work, we have presented a unified model that combines several state-of-the-art models, in particular BERT for text representation and MolBERT for drug representation. Our models led to significant improvements on ADE classification of French texts (by 8\% with the previous state-of-the-art quality of 17\% F1) and achieved state-of-the-art results on recent SMM4H 2021 Task 1a and Task 2 for English and Russian texts, respectively, all of this using only well-known BERT-based pre-trained models for individual components. Possible directions for future work include: (i) multilingual learning on several datasets; (ii) multi-task learning on two tasks: ADE text classification and prediction of the side effects of a given drug using drug structure or knowledge graphs with interactions between biomedical entities; (iii) zero-shot cross-drug class evaluation across drug classes.

%% file: data.tex

\begin{table}[!t]
\centering
\caption{Summary of statistics of experimental datasets. \label{tab:stats}} 
\begin{tabular}{|l|l|l|l|l|}
\hline
 & \textbf{Train} & \textbf{Dev} & \textbf{Test}  & \textbf{All} \\ \hline
\multicolumn{5}{|c|}{SMM4H 2020 Task 1b, French tweets} \\ \hline
\# tweets & 1,941 (1.6\%) & 485 (1.6\%) & 607 (1.6\%) & 3,033 (1.6\%)  \\ 
\#  tweets with ADEs & 31  & 8  & 10 & 49 \\
\# unique drug ids & 175  & 91 & 114 & 219 \\ 
avg. len & 27.54 & 26.72 & 27.45 & 27.39 \\ \hline
\multicolumn{5}{|c|}{SMM4H 2021 Task 2, Russian tweets} \\ \hline
\# tweets & 10,609 & 1,000 & 9,095 & 20,707 \\ 
\#  tweets with ADEs & 980 (9.2\%) & 92 (9.2\%) & 778 (8.6\%) & 1,850 (8.9\%) \\ 
\# unique drug ids & 62  & 47 & 41 & 65 \\
avg. len & 21.48 & 22.51 & 18.65 & 20.29 \\ \hline
\multicolumn{5}{|c|}{SMM4H 2021 Task 1a, English tweets} \\ \hline
\# tweets &  13,908 & 914  & 3,477 & 18,299   \\ 
\#  tweets with ADEs & 988 (7.1\%) & 65 (7.1\%) & 247 (7.1\%) & 1,300 (7.1\%)  \\  
\# unique drug ids & 581 & 215 & 340 & 638 \\ 
avg. len & 19.81 & 19.75 & 19.72 & 19.79 \\ \hline

\multicolumn{5}{|c|}{SMM4H 2021 Task 1a, English tweets, ATC-stratified split} \\ \hline

\# tweets & 13,176  & 1,456 & 3,667 & 18,299   \\ 
\#  tweets with ADEs & 968 (7.3\%) & 102 (7\%)  & 230 (6.3\%) & 1,300 (7.1\%)   \\  
\# unique drug ids & 582  & 234  & 349  & 638  \\ 
avg. len & 19.74 & 20.03 & 19.89 & 19.79 \\ \hline
\end{tabular}
\end{table}

\begin{table}[!t]
\centering
\caption{Most frequent drugs of SMM4H corpora. The number in parenthesis indicates the number of tweets with the drug mention. \label{tab:top_drugs}} 
\begin{tabular}{|l|l|}
\hline      
\textbf{Dataset} & \textbf{Most frequent drugs with DrugBank id} \\\hline 
\vtop{\hbox{\strut SMM4H 2020 Task 1b,}\hbox{\strut French tweets}} & \vtop{\hbox{\strut xanax, DB00404 (929)}\hbox{\strut ventoline, DB01001 (606)}\hbox{\strut prozac, DB00472 (226)}\hbox{\strut levothyrox, DB00451 (203)}\hbox{\strut morphine, DB00295 (101)}} \\\hline
\vtop{\hbox{\strut SMM4H 2021 Task 2,}\hbox{\strut Russian tweets}} & \vtop{\hbox{\strut paracetamol, DB00316 (6124)}\hbox{\strut umifenovir/arbidol, DB13609 (2363)}\hbox{\strut prozac, DB00472 (2335)}\hbox{\strut xanax, DB00404 (1718)}\hbox{\strut paroxetine, DB00715 (834)}} \\\hline
\vtop{\hbox{\strut SMM4H 2021 Task 1a,}\hbox{\strut English tweets}} & \vtop{\hbox{\strut adderall, DB01576 (1545)}\hbox{\strut prozac, DB00472 (1109)}\hbox{\strut vyvanse, DB01255 (900)}\hbox{\strut cymbalta, DB00476 (863)}\hbox{\strut xanax, DB00404 (814)}} \\\hline
\end{tabular}{}
\end{table}




The ADE classification task involves distinguishing tweets that report an adverse effect of a medication (annotated as ``ADE") from those that do not (annotated as ``non-ADE"). 
The aim of the Social Media Mining for Health Applications (\#SMM4H) shared tasks is to take a community-driven approach to addressing NLP challenges of utilizing social media data for health informatics, including informal, colloquial expressions of clinical concepts, noise, data sparsity, ambiguity, and multilingual posts. In 2020, the fifth iteration of the SMM4H shared tasks included Task 2 on automatic classification of multilingual tweets that report adverse effects of a medication. This dataset includes tweets posted in English, French, and Russian \citep{klein2020overview}. 

All SMM4H datasets are manually labeled using the same annotation guidelines. According to these guidelines, an ADE or ADR is an effect of the drug that is not desired. This includes a worsening of the indication, drug withdrawal and its associated symptoms, and a loss of effectiveness of the medication. The general rules for marking a tweet as positive for containing a mention of an ADR are:
\begin{enumerate*}
\item The user states an ADR that resulted from taking the drug;
\item The ADR mentioned had been personally experienced by the user, or by someone known personally to the user;
\item The ADR mentioned is specific, that is it refers to a sign or symptom that can be coded to a medical term even if the mention itself is in colloquial terms;
\item The ADR occurred over the normal course of treatment and is not an effect from taking too much of the drug;
\item The ADR is not the result of taking two, or more, drugs at the same time (drug-drug interactions).
\end{enumerate*}

English-language version of the SMM4H task is a long-running task starting from 2016, while French and Russian subtasks were presented for the first time. Within a year, \cite{magge2021overview} released English and Russian datasets for SMM4H 2021 Task 1a and Task 2, respectively. We note that all Russian texts from SMM4H 2020 were included in the SMM4H 2021 Task 2 train/dev sets. Therefore, we use the latter one for our experiments. Since the organizers of the SMM4H 2021 Shared Task did not publish the original annotated test set, we use 20\% of the original train tweets as a test set for our experiments. The data split was performed randomly, but the positive and negative class proportion is kept the same for train and test sets. To explore the effect of drug group distribution among train and test sets, we created an additional ATC-stratified split of the English SMM4H 2021 dataset. In the ATC-stratified data split, the proportion of tweets mentioning an ATC group is the same for train and test sets for a fixed ATC group.

Table \ref{tab:stats} presents various statistics of these subsets. 
Several interesting observations can be made based on Table \ref{tab:stats}. First, the size of training sets varies significantly: the training set for the English subtask contains 25,678 tweets, while the first-running French set contains only 1,941 tweets. Second, the task is especially challenging due to the imbalanced nature of the dataset where the tweets that mention ADE are outnumbered 10:1 to 50:1 by tweets that do not contain ADEs. Third, the number of unique drugs differs significantly for different languages. The definition of drug products and dietary supplements provided by the FDA \citep{FDA20} were used to collect English tweets. The French tweets have been selected based on a list of around 120 drug names. The Russian tweets were collected as follows. For the Russian subset for SMM4H 2020, around 70 drugs most commonly described in English tweets were manually associated with drugs on the Russian market using the State Register of Medicinal Remedies \citep{GRLS21}. Additionally, the authors collected tweets for COVID-related drugs such as \textit{paracetamol} and \textit{unifemovir} in 2021 for the test set. Paracetamol, also known as acetaminophen, is used to treat fever and mild pain. Umifenovir, also known as arbidol, is a popular antiviral medication for the treatment of influenza infection used in Russia. 

\begin{figure}[!t]
\centerline{\includegraphics[width=1\linewidth]{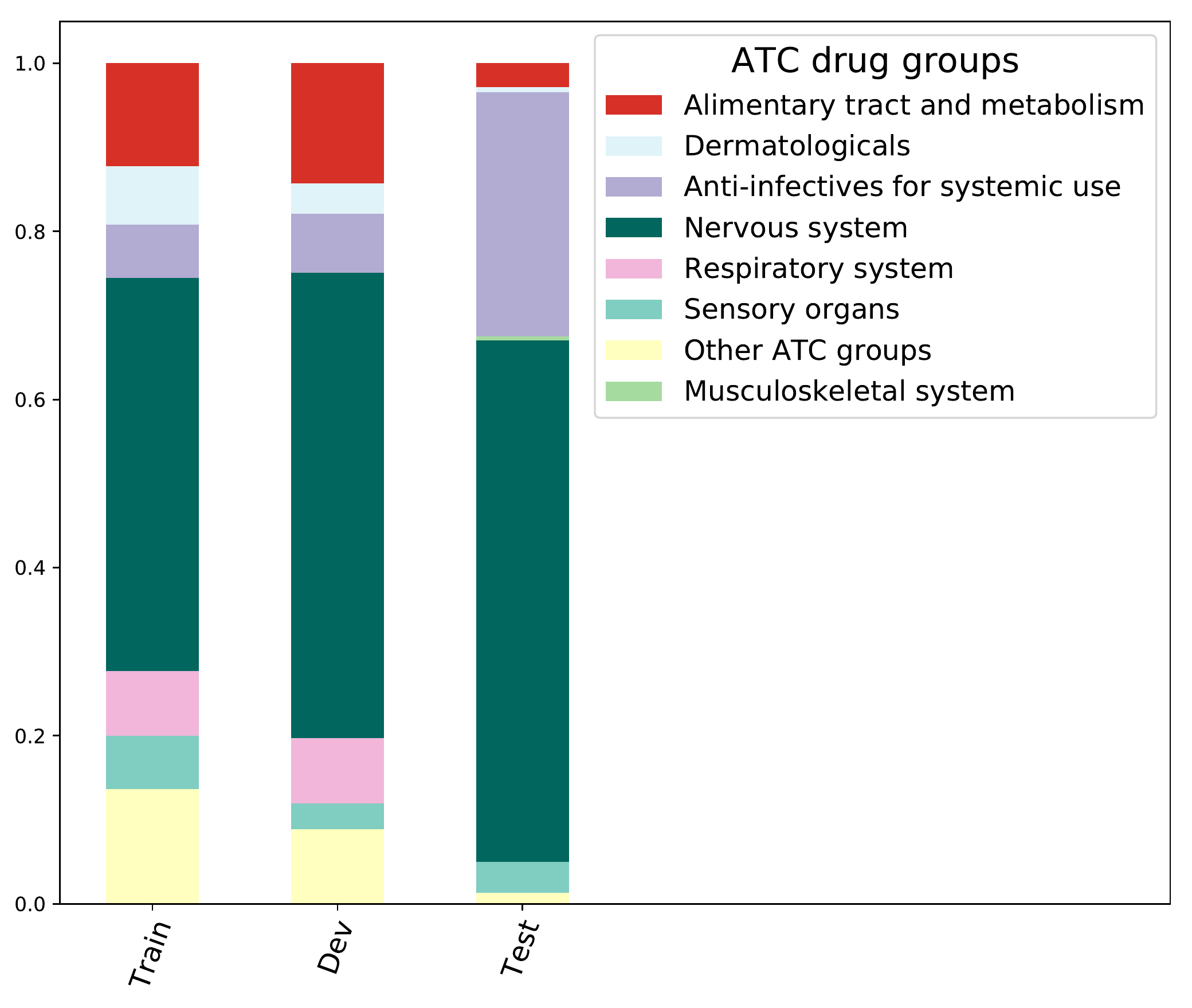}}
\caption{ATC classes distribution in the Russian corpus for SMM4H 2021.}\label{fig:stats}
\end{figure}

Table \ref{tab:top_drugs} shows most frequent drugs of SMM4H corpora. Top $5$ drugs in English tweets are parts of the ATC group \textit{Nervous system} (N). Russian tweets include drugs assigned to two groups: \textit{Antiinfectives for systemic use} (J) and \textit{Nervous system} (N). French tweets include drugs assigned to four groups: \textit{Nervous system} (N), \textit{Respiratory system} (R), \textit{Systemic hormonal preparations, excluding sex hormones and insulins} (H), \textit{Alimentary tract and metabolism} (A). 
We analyzed the number of tweets associated with drugs from different ATC classes for the Russian subset for SMM4H 2021. Figure \ref{fig:stats} presents these results. As expected, some groups such as \textit{Antiinfectives for systemic use} are represented in the test set more significantly than in the train set.

\begin{figure}
     \centering
     \begin{subfigure}[b]{0.4\textwidth}
         \centering
         \includegraphics[width=\textwidth]{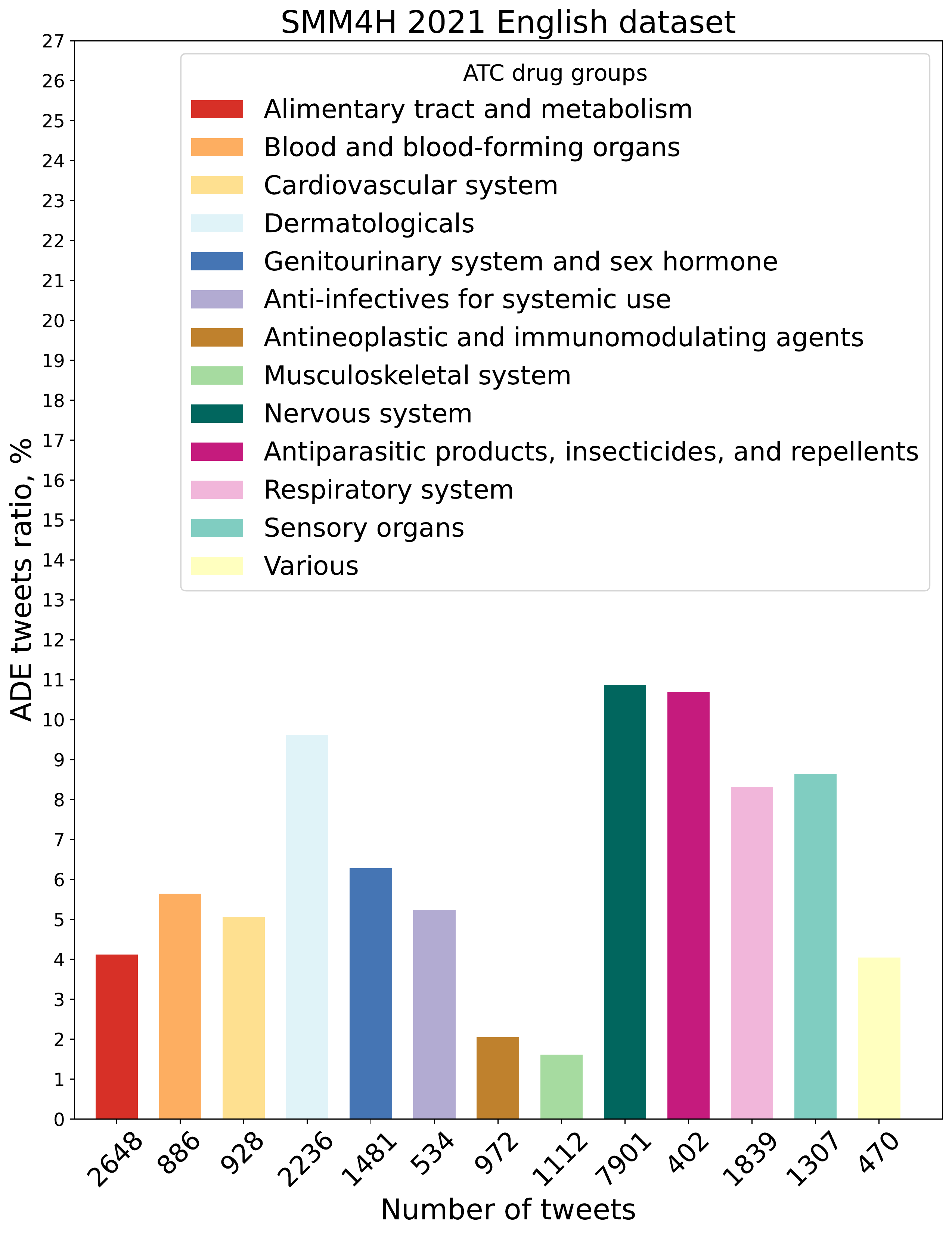}
     \end{subfigure}
     \hfill
     \begin{subfigure}[b]{0.4\textwidth}
         \centering
         \includegraphics[width=\textwidth]{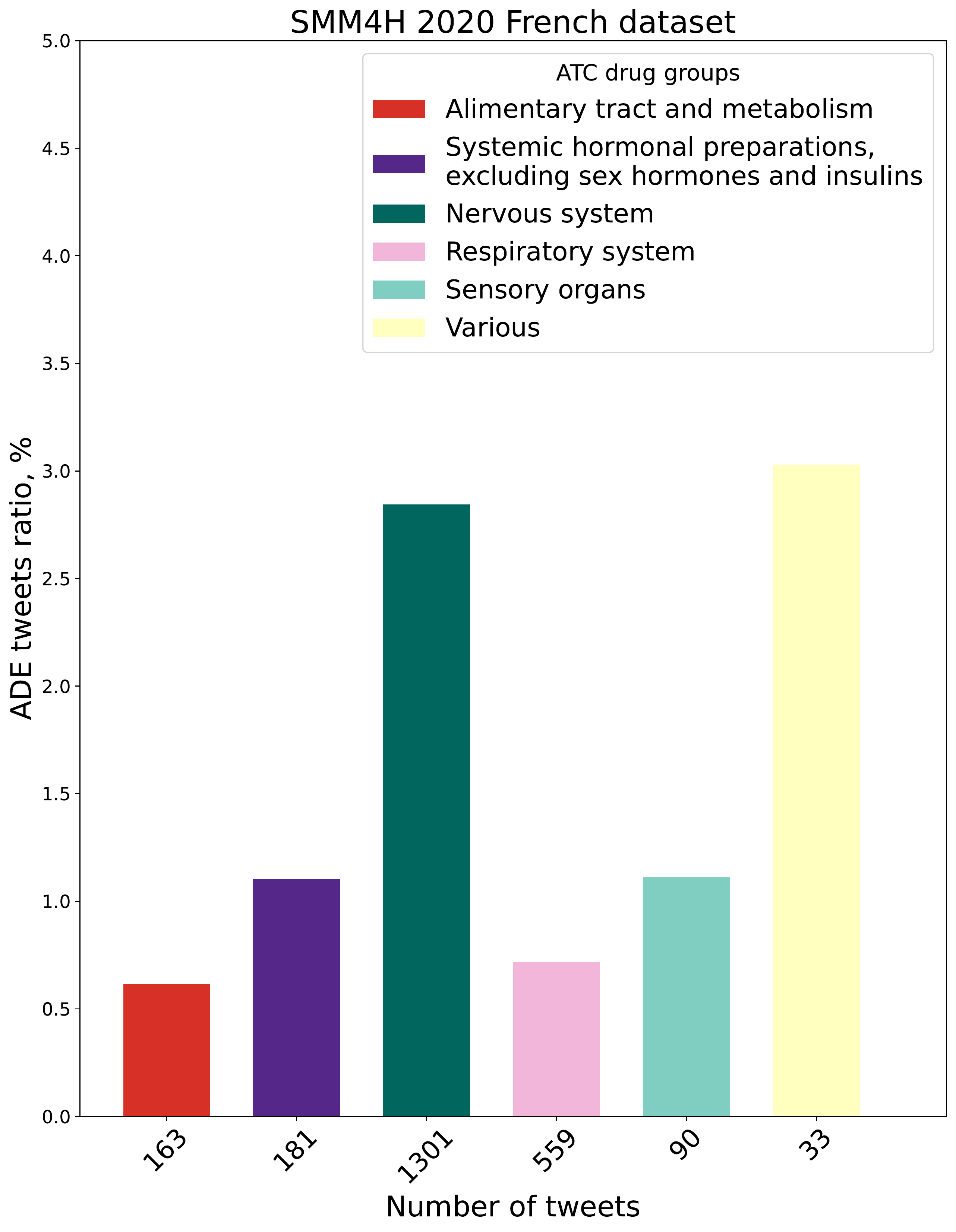}
     \end{subfigure}
     \hfill
     \begin{subfigure}[b]{0.4\textwidth}
         \centering
         \includegraphics[width=\textwidth]{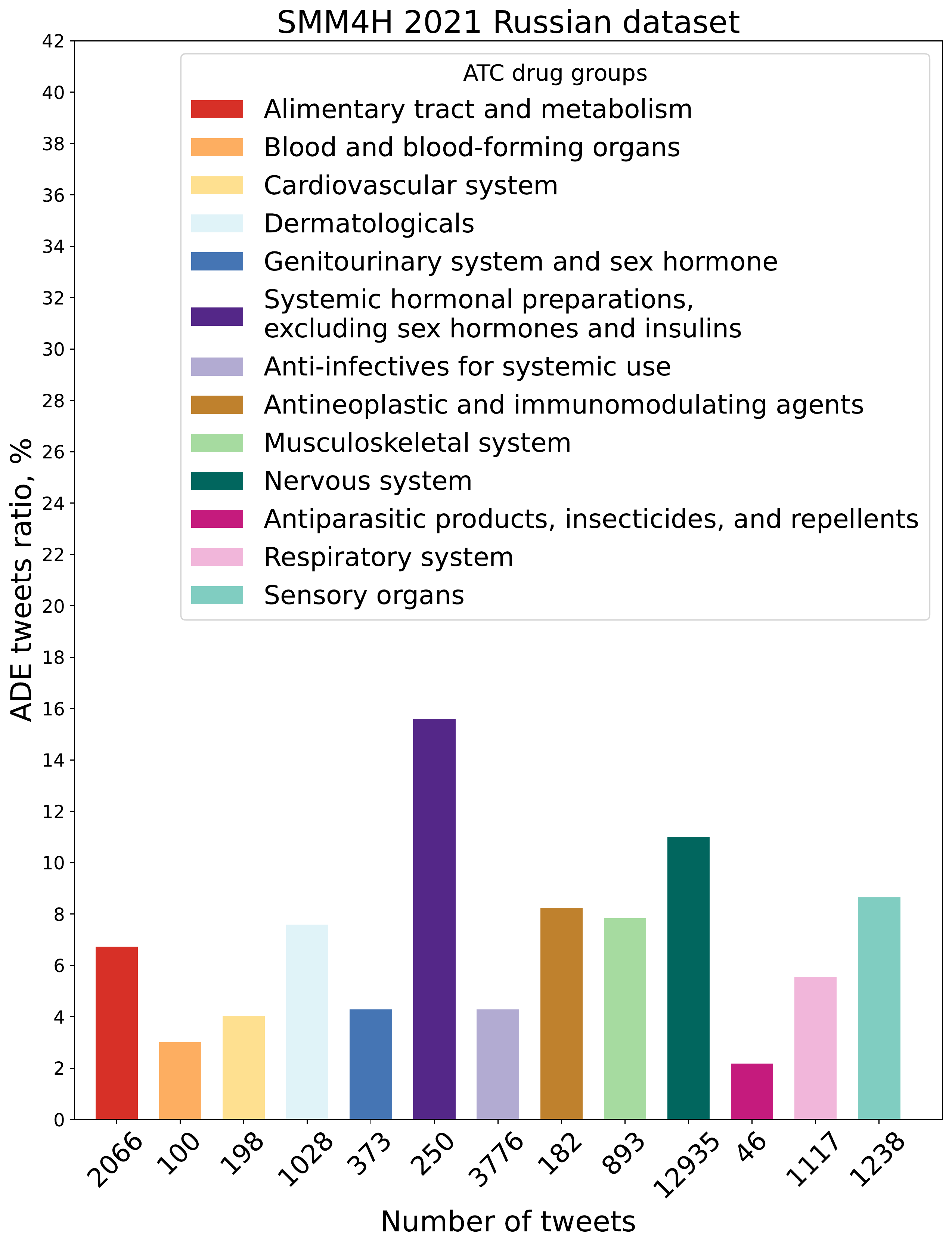}
     \end{subfigure}
        \caption{Percentage of tweets containing ADE mentions grouped by ATC group for different SMM4H corpora. ATC groups with no positive class samples are not shown.}
        \label{fig:ade_ratio_by_atc_group}
\end{figure}

To determine whether an ATC group of a drug mentioned in a text provides an apriori knowledge on how probable is the text to report an ADE, we calculated the statistics on the percentage of ADE tweets among tweets of each ATC class. The results are presented in figure~\ref{fig:ade_ratio_by_atc_group}. Several important observations have to be done. First, the percentage of tweets mentioning ADE varies significantly among different drug groups. For example, about 2\% of English tweets of the "\textit{Antineoplastic and immunomodulating agents}" class mention an adverse effect, whereas more than 10\% tweets of the "\textit{Nervous systems}" class mention an ADE. Thus, we can conclude that given a text with drug mention, we are provided with some apriori knowledge on how likely is the text to report an ADE. Second, the proportion of tweets mentioning an ADE is inconsistent among different languages, even for the same ATC group. For example, for "\textit{Musculoskeletal system}" class about 7\% of Russian tweets and less than 2\% of English tweets report an ADE. Based on the observation, we can hypothesize that drug-related knowledge transfer across multiple languages may fail due to this inconsistency. Besides, drug-related information is not an exhaustive tweet representation itself, and textual context is required.

For our experiments, we use the French sets from the SMM4H 2020 task and English and Russian sets from the SMM4H 2021 tasks. The data preprocessing pipeline is  identical for all datasets and is adopted from~\citep{miftahutdinov2020kfu} and includes the following steps: (i) replacement of all URLs with the word ``link''; (ii) masking of user mentions with the ``@username'' placeholder; (iii) replacement of some emojis with a textual representation (e.g., pill and syringe emojis are replaced by ``pill'' and ``syringe'' words).



%% file: tables.tex
\begin{table}[!t]
\centering
\caption{Evaluation results for SMM4H 2020 Task 1b on classification of French tweets. “$\dagger$” denotes statistically significantly better than unimodal BERT (Wilcoxon, $\rho < 0.01$).}
\label{tab:smm4h_20_fr_results}
\begin{tabular}{|l|c|c|c|}
\hline
\textbf{Model} & \textbf{P $\pm$ std} & \textbf{R $\pm$ std} & \textbf{F\textsubscript{1} $\pm$ std} \\ \hline
\multicolumn{4}{|c|}{Text-only models} \\ \hline
TF-IDF + SVM  & 0.769   & 0.2  & 0.111 \\ 
Fasttext + CNN  & 0.018   & 0.7  & 0.034 \\ 
BERT      & 0.196 $\pm$ 0.170   & 0.2 $\pm$ 0.133   & 0.169 $\pm$ 0.108 \\ \hline
\cite{gencoglu2020sentence}    & 0.15 & 0.20 & 0.17 \\  
BERT      & 0.196 $\pm$ 0.170   & 0.2 $\pm$ 0.133   & 0.169 $\pm$ 0.108 \\ \hline
\multicolumn{4}{|c|}{Models with concatenated modalities} \\ \hline
BERT + ATC categories     & 0.185 $\pm$ 0.049  & 0.23 $\pm$ 0.134  & 0.190 $\pm$ 0.063  \\ 
BERT + descriptors     & 0.317 $\pm$ 0.176 & 0.22 $\pm$ 0.092  & 0.245 $\pm$ 0.111$^\dagger$  \\ 
BERT + ChemBERTa        & 0.204 $\pm$ 0.221  & 0.14 $\pm$ 0.135  &  0.154 $\pm$ 0.151  \\ 
BERT + MolBERT       & 0.415 $\pm$ 0.233   & 0.23 $\pm$ 0.106 & \textbf{0.254 $\pm$ 0.071}$^\dagger$ \\ \hline
\multicolumn{4}{|c|}{Models with cross-attention} \\ \hline
BERT + ATC categories     & 0.249 $\pm$ 0.302 & 0.09 $\pm$ 0.057  & 0.113 $\pm$ 0.069  \\ 
BERT + descriptors     &  0.260 $\pm$ 0.199 & 0.14 $\pm$ 0.084  & 0.164 $\pm$ 0.086  \\ 
BERT + ChemBERTa        & 0.225 $\pm$ 0.083 & 0.17 $\pm$ 0.082   & 0.187 $\pm$ 0.075   \\ 
BERT + MolBERT       & 0.251 $\pm$ 0.142  & 0.16 $\pm$ 0.117  &  0.178 $\pm$ 0.099  \\ \hline
\end{tabular}
\end{table}

\begin{table}[!t]
\centering
\caption{Evaluation results for SMM4H 2021 Task 2 on classification of Russian tweets \citep{magge2021overview}. We did not observe a statistically significant (Wilcoxon, $\rho < 0.05$) difference between bi-modal models and unimodal BERT. \label{tab:smm4h_21_ru_results}}
\begin{tabular}{|p{3.73cm}|c|c|c|}
\hline
\textbf{Model} & \textbf{P $\pm$ std} & \textbf{R $\pm$ std} & \textbf{F\textsubscript{1} $\pm$ std} \\ \hline
\multicolumn{4}{|c|}{Text-only models} \\ \hline
TF-IDF + SVM  & 0.158 & 0.281  & 0.202 \\ 
Fasttext + CNN  & 0.356   & 0.465  & 0.404 \\ 
BERT      & 0.548 $\pm$ 0.063  & 0.516 $\pm$ 0.072  & 0.524 $\pm$ 0.020  \\ \hline
\multicolumn{4}{|c|}{Models with concatenated modalities} \\ \hline
BERT + ATC categories     & 0.529 $\pm$ 0.054  & 0.521 $\pm$ 0.058  & 0.519 $\pm$ 0.090   \\ 
BERT + descriptors     & 0.543 $\pm$ 0.052   & 0.514  $\pm$ 0.059  & 0.523 $\pm$ 0.020   \\ 
BERT + ChemBERTa        & 0.552 $\pm$ 0.061  & 0.513 $\pm$ 0.076  & 0.524 $\pm$ 0.023   \\ 
BERT + MolBERT       &  0.538 $\pm$ 0.040  &  0.531 $\pm$ 0.040  & \textbf{0.532 $\pm$ 0.014}   \\ \hline
\multicolumn{4}{|c|}{Models with cross-attention} \\ \hline
BERT + ATC categories     & 0.509 $\pm$ 0.063  & 0.542 $\pm$ 0.059  & 0.519  $\pm$ 0.011 \\ 
BERT + descriptors     & 0.527 $\pm$ 0.076  & 0.518 $\pm$ 0.073  & 0.514 $\pm$ 0.022  \\ 
BERT + ChemBERTa        & 0.515  $\pm$ 0.049  & 0.537  $\pm$ 0.063  & 0.521  $\pm$ 0.020   \\ 
BERT + MolBERT       & 0.514 $\pm$ 0.051  & 0.553 $\pm$ 0.055 & 0.529 $\pm$ 0.011 \\ \hline
\end{tabular}
\end{table}


\begin{table}[!t]
\centering
\caption{Evaluation results for on classification of English tweets on SMM4H 2021 dataset (unofficial class-stratified test set). “$\dagger$” denotes statistically significantly better than unimodal RoBERTa (Wilcoxon, $ \rho < 0.01$). \label{tab:smm4h_21_en_results}}
\begin{tabular}{|p{3.7cm}|c|c|c|}
\hline
\textbf{Model} & \textbf{P $\pm$ std} & \textbf{R $\pm$ std} & \textbf{F\textsubscript{1} $\pm$ std} \\ 
\hline
\multicolumn{4}{|c|}{Text-only models} \\ \hline
TF-IDF + SVM  & 0.484   & 0.417  & 0.448 \\ 
Word2vec + CNN  & 0.664   & 0.344  & 0.453 \\ 
RoBERTa       & 0.755 $\pm$ 0.062   & 0.842 $\pm$ 0.051 & 0.793 $\pm$ 0.022   \\ \hline
Team 4 \citep{ramesh2021bert}       &  0.59  & 0.92  & 0.72   \\ \hline
\multicolumn{4}{|c|}{Models with concatenated modalities} \\ \hline
RoBERTa + ATC categories     & 0.761 $\pm$ 0.048   & 0.829 $\pm$ 0.064   & 0.791 $\pm$ 0.025    \\ 
RoBERTa + descriptors     & 0.737 $\pm$ 0.045   & 0.838 $\pm$ 0.051   & 0.782 $\pm$ 0.016    \\ 
RoBERTa + ChemBERTa        & 0.757 $\pm$ 0.055  & 0.828 $\pm$ 0.057  & 0.787 $\pm$ 0.014   \\ 
RoBERTa + MolBERT       & 0.762 $\pm$ 0.044  & 0.829 $\pm$ 0.057   & 0.792 $\pm$ 0.019   \\ \hline
\multicolumn{4}{|c|}{Models with cross-attentions} \\ \hline
RoBERTa + ATC categories     & 0.770 $\pm$ 0.051   & 0.812 $\pm$ 0.058  & 0.788 $\pm$ 0.022   \\ 
RoBERTa + descriptors     & 0.794 $\pm$ 0.031  & 0.808 $\pm$ 0.047   & \textbf{0.799 $\pm$ 0.021$^\dagger$}  \\ 
RoBERTa + ChemBERTa        & 0.771 $\pm$ 0.065  & 0.822 $\pm$ 0.054  & 0.792 $\pm$ 0.025  \\ 
RoBERTa + MolBERT   & 0.760 $\pm$ 0.057  & 0.832 $\pm$ 0.034 & 0.793 $\pm$ 0.031  \\
\hline
\end{tabular}
\end{table}

\begin{table}[!t]
\centering
\caption{Official SMM4H 2021 results \citep{magge2021overview} on the official test set.  \label{tab:smm4h_21_ru_results_magge}}
\begin{tabular}{|c|c|c|c|}
\hline
\textbf{Model} & \textbf{P} & \textbf{R} & \textbf{F\textsubscript{1}} \\ \hline
\multicolumn{4}{|c|}{Task 2, Russian} \\ \hline
Our SMM4H 2021 Task 2 submission \citep{sakhovskiy2021kfu} & 0.58   & 0.57 & \textbf{0.57} \\ 
Second place \citep{magge2021overview} & 0.54 & 0.57 & 0.52 \\
\hline
\multicolumn{4}{|c|}{Task 1a, English} \\ \hline
Our SMM4H 2021 Task 1a submission \citep{sakhovskiy2021kfu} & 0.552 & 0.681 & \textbf{0.61} \\ 
First/second place, Team 4 \citep{ramesh2021bert}
 & 0.515 & 0.752 & \textbf{0.61} \\
\hline
\end{tabular}
\end{table}

\begin{table}[!t]
\centering
\caption{Evaluation results for  classification of English tweets on the ATC-stratified SMM4H 2021 test set. “$\dagger$” denotes statistically significantly better than unimodal RoBERTa (Wilcoxon, $ \rho < 0.05$). \label{tab:smm4h_21_stratified_en_results}}
\begin{tabular}{|p{3.7cm}|c|c|c|}
\hline
\textbf{Model} & \textbf{P $\pm$ std} & \textbf{R $\pm$ std} & \textbf{F\textsubscript{1} $\pm$ std} \\ 
\hline

\multicolumn{4}{|c|}{Text-only models} \\ \hline
RoBERTa       & 0.682 $\pm$ 0.058  & 0.764 $\pm$ 0.076 & 0.715 $\pm$ 0.032  \\ \hline
\multicolumn{4}{|c|}{Models with concatenated modalities} \\ \hline
RoBERTa + ATC categories     & 0.675 $\pm$ 0.040  & 0.782 $\pm$ 0.038   & 0.723 $\pm$ 0.019    \\ 
RoBERTa + descriptors     & 0.679 $\pm$ 0.059   & 0.786 $\pm$ 0.044  & 0.725 $\pm$ $0.019^{\dagger}$   \\ 
RoBERTa + ChemBERTa        & 0.685 $\pm$ 0.049 & 0.778 $\pm$ 0.037  & \textbf{0.726 $\pm$ 0.014}  \\ 
RoBERTa + MolBERT       & 0.654 $\pm$ 0.055  & 0.792 $\pm$ 0.051  & 0.713 $\pm$ 0.018  \\ \hline
\multicolumn{4}{|c|}{Models with cross-attentions} \\ \hline
RoBERTa + ATC categories     & 0.682 $\pm$ 0.063   & 0.762 $\pm$ 0.057  & 0.716 $\pm$ 0.022   \\
RoBERTa + descriptors     & 0.684 $\pm$ 0.062   & 0.762 $\pm$  0.051  & 0.717 $\pm$ 0.017  \\ 
RoBERTa + ChemBERTa        & 0.676 $\pm$ 0.061 & 0.782 $\pm$ 0.055  & 0.721 $\pm$ 0.020  \\ 
RoBERTa + MolBERT   & 0.668 $\pm$ 0.070  & 0.784 $\pm$ 0.053 & 0.717 $\pm$ 0.030  \\

\hline
\end{tabular}
\end{table}

%% file: tables2.tex
\begin{table}[h!t]
\centering
\caption{Comparison of uni- and bi-modal models performance on various therapeutic groups. The number in parenthesis indicates the number of tweets about drugs associated with the ATC class.  \label{tab:quality_by_group}} 
\begin{tabular}{|p{3.85cm}|c|c|c|c|c|c|c|}
\hline
\multirow{2}{*}{\textbf{\shortstack{ATC group}}}
 &  \multicolumn{3}{|c|}{\textbf{Uni-modal model}}  & \multicolumn{3}{|c|}{\textbf{Bi-modal model}}  \\ \cline{2-7}
& \textbf{P} & \textbf{R}  & \textbf{F\textsubscript{1}} & \textbf{P} & \textbf{R}  & \textbf{F\textsubscript{1} } \\ \hline

\multicolumn{7}{|c|}{\textbf{SMM4H 2021 Task 2, Russian tweets, official test set}} \\ \hline
Nervous system (6,100) & 0.456 & 0.562 & 0.495 &  0.472 & 0.558 & 0.506 \\ 
Anti-infectives for systemic use (2,859) & 0.440 & 0.505 & 0.464 & 0.448 & 0.495 & 0.465 \\
Sensory organs (365) & 0.767 & 0.720 & 0.738 & 0.805 & 0.747 & 0.771 \\
Alimentary tract and metabolism (280) & 0.675 & 0.669 & 0.662 & 0.713 & 0.652 & 0.670 \\ 
Dermatologicals (61) & 0.967 & 0.725 & 0.825 & 0.974 & 0.75 & 0.845 \\\hline

\multicolumn{7}{|c|}{\textbf{SMM4H 2021 Task 1a, English tweets, official dev set}} \\ \hline
Nervous system (470) & 0.734 &  0.864 & 0.791 & 0.785 & 0.830 & 0.805 \\
Dermatologicals (158) & 0.613 & 0.842 & 0.703 & 0.626 & 0.808 & 0.705 \\
Alimentary tract and metabolism (155) & 0.767 & 0.85 & 0.804 & 0.816 & 0.75 & 0.774  \\
Respiratory system (130) & 0.658 & 0.925 & 0.760 & 0.678 & 0.875 & 0.760 \\
Sensory organs (88) & 0.832 & 0.880 & 0.851 & 0.849 & 0.79 & 0.814 \\\hline

\hline

\end{tabular}
\end{table}